\documentstyle[amssymb,twocolumn,aps,prb]{revtex}

\begin{document}
\title{{\bf Band- and k-dependent Self-Energy Effects in }\\
{\bf the Unoccupied and Occupied Quasiparticle Band Structure of Cu}}
\author{V.N. Strocov, R. Claessen}
\address{{Experimentalphysik II, Universit\"{a}t Augsburg, D-86135 }%
Augsburg, Germany }
\author{F. Aryasetiawan}
\address{Research Institute for Computational Sciences, AIST, Tsukuba
Central 2, 1-1-1 Umezono, Tsukuba, Ibaraki 305-8568, Japan }
\author{P. Blaha}
\address{Institut f\"{u}r Physikalische und Theoretische Chemie, Technische
Universit\"{a}t Wien, A-1060 Wien, Austria}
\author{P.O. Nilsson}
\address{Department of Physics, Chalmers University of Technology, SE-41296 G%
\"{o}teborg, Sweden}
\date{\today }
\maketitle

\begin{abstract}
Excited-state self-energy effects in the electronic structure of Cu, a
prototype weakly correlated system containing states with different degrees
of localization, are investigated with emphasis on the unoccupied states up
to 40 eV above the Fermi level. The analysis employs the experimental
quasiparticle states mapped under full control of the 3-dimensional
wavevector ${\bf k}$ using very-low energy electron diffraction for the
unoccupied states and photoemission for the occupied states. The self-energy
corrections to the density-functional theory show a distinct band- and ${\bf %
k}$-dependence. This is supported by quasiparticle {\it GW} calculations
performed within the framework of linearized muffin-tin orbitals. Our
results suggest however that the {\it GW} approximation may be less accurate
in the localized {\it d}-bands of Cu with their short-range charge
fluctuations. We identify a connection of the self-energy behavior with the
spatial localization of the one-electron wavefunctions in the unit cell and
with their behavior in the core region. Mechanisms of this connection are
discussed based on the local-density picture and on the non-local exchange
interaction with the valence states.
\end{abstract}


\section{Introduction}

Many-body exchange-correlation effects in the inhomogeneous interacting
electron system of solids, reflected by the quasiparticle $E({\bf k})$ band
structure observed in the experiment, are still far from complete
understanding. The standard Density Functional Theory (DFT) accounts for
these effects only in the ground state, yielding the static properties of
the solids such as the charge density or the total energy. Of the DFT\
eigenvalues $\varepsilon ({\bf k})$, however, only the highest occupied one
yields the correct excitation energy by a DFT analogue to the Koopmans
theorem,\cite{Koopmans} while the others, strictly speaking, have no
physical meaning as energy levels or excitation energies. Description of the
quasiparticle excited states, created by external photon, electron, etc.
impact, is by far more difficult (see, e.g., [\ref{Hedin69},\ref{Kevan92}]).
Their energies $E({\bf k})$ deviate from $\varepsilon ({\bf k})$ owing to
the difference of the excited-state {\em dynamic} exchange-correlation
potential, or the self-energy, from the ground-state {\em static}
exchange-correlation potential $V_{XC}$ used in the DFT. The former is
described by a complex energy- and ${\bf k}$-dependent non-local self-energy
operator ${\bf \Sigma }$, whose imaginary part $Im\Sigma $ describes the
lifetime broadening of the excitation. Whereas for materials with strong
local on-site correlations the excited-state many-body effects are usually
described by model Hamiltonians, this approach cannot be extended to weakly
correlated materials with long-range charge fluctuations. Only a few
perturbational low-order approaches, such as the $GW$ approximation,\cite%
{Hedin69,Hedin99,Aryasetiawan98} are suggested for this case.

Cu has since long been believed to be a prototype weakly-correlated metal
for which the DFT ground-state picture provides an accurate description of
the excitation spectra (see, e.g., the review [\ref{Courths84}]). Only
recently, based on comparison of new angle-resolved photoemission (PE) data
obtained under full control of the 3-dimensional wavevector ${\bf k}$ with
state-of-art DFT calculations, we have demonstrated that the self-energy
corrections $\Delta \Sigma =E({\bf k})-\varepsilon ({\bf k})$ to the DFT\ in
the valence band of Cu display a clear ${\bf k}$- and band dependence,
reaching values as large as 0.5 eV in the {\it d}-bands.\cite{CFS} These
results gave a serious indication for the importance of self-energy effects
even in such a supposedly simple metal as Cu. In fact, the previous
misconception had arisen only due to errors of the early DFT\ calculations
which accidentally matched $\Delta \Sigma $.\cite{Courths01} Our preliminary 
$GW$ calculations\cite{GW-SRL} have confirmed the general trends of the
observed self-energy behavior. The most recent $GW$ calculations\cite%
{Marini02} performed within the pseudopotential framework have yielded a
striking agreement with the experimental valence band of Cu concerning the $%
{\bf k}$- and band dependence of the $\Delta \Sigma $ shifts as well as
their absolute values.

While the self-energy effects in the valence band can be studied using PE
spectroscopy, the unoccupied states still remain largely unexplored.
Conventional spectroscopies of the upper states using X-ray absorption or
Bremsstrahlung phenomena\cite{Fuggle92} provide only ${\bf k}$-integrated
information such as the matrix element weighted density of states. The ${\bf %
k}$-resolving techniques such as inverse PE still involve two electron
states and suffer from uncertainties in the surface-perpendicular wavevector 
$k_{\perp }$. Only recently it was realized that in the energy region above
the vacuum level the quasiparticle $E({\bf k})$ can be mapped under full
control of the 3-dimensional ${\bf k}$ using Very-Low-Energy Electron
Diffraction (VLEED) (see [\ref{CFS},\ref{Exi},\ref{TMDCs}] and references
therein). As the PE final states are the time-reversed LEED states,\cite%
{Feibelman74} this information can then be used in PE spectroscopy to
achieve 3-dimensional mapping of the valence band $E({\bf k})$.\cite%
{CFS,TMDCs}

Here, we extend the self-energy analysis to the unoccupied states in the
energy region from the vacuum level up to 40 eV above the Fermi level $E_{F}$
based on the VLEED experimental data. Contrary to a commonly accepted view,
in this energy region the $\Delta \Sigma $ shifts also show a clear ${\bf k}$%
- and band dependence. We support our findings by $GW$ calculations
performed within the framework of linearized muffin-tin orbitals, which
include the valence and core states on equal footing. We show that the $%
\Delta \Sigma $ anomalies correlate with the spatial character of the
one-electron wavefunctions $\phi _{{\bf k}}$, and endeavour to identify the
essential physics of this. Our results on the valence band suggest that the $%
GW$ approximation underestimates the self-energy shifts due to the
short-range charge fluctuations characteristic of the {\it d}-bands of Cu.

\section{Experimental procedure and results}

Our analysis employs the results of the recent VLEED\ and PE measurements
performed on the Cu(110) surface.\cite{CFS} The information about the bulk
band structure, reflected in the VLEED spectra, was insensitive to a slight
oscillatory relaxation of this surface.\cite{Relax} This was confirmed by
measurements on different surfaces\cite{CFS} as well as dynamical VLEED
calculations.\cite{FederCalc}

Mapping of the unoccupied quasiparticle $E({\bf k})$ using VLEED (see [\ref%
{CFS},\ref{Exi},\ref{TMDCs}] and the references therein) is based on the
fact that the diffraction process is connected with $E(k_{\bot })$ along the
Brillouin zone (BZ) direction determined by parallel momentum conservation
(external ${\bf K}_{\parallel }$ = internal ${\bf k}_{\parallel }$). In
particular, the extrema in the differential VLEED spectrum $dT/dE$ of the
elastic electron transmission reflect the critical points (CPs) in the $%
k_{\bot }$ dispersions of the bands whose Bloch wave $\phi _{{\bf k}}$
efficiently couples to the incident plane wave. ${\bf K}_{\parallel }$
dispersion of the $dT/dE$ extrema, reflecting the CPs with $k_{\bot }$ lying
on a surface-parallel symmetry line of the BZ, yields then $E({\bf k})$
along this line resolved in the 3-dimensional ${\bf k}$. In our case the
angle-dependent VLEED data with ${\bf K}_{\parallel }$ varying in the $%
\overline{\Gamma }\overline{Y}$ azimuth of the surface BZ yielded $E({\bf k}%
) $ along the $\Gamma X$ line of the bulk BZ.\cite{CFS}

The direct mapping of the $dT/dE$ extrema reflects however the CPs in the
quasiparticle $E({\bf k})$ {\em near the surface}, corresponding to $\phi _{%
{\bf k}}$ excited by the VLEED\ beam and damped towards the crystal interior
due to the finite electron lifetime, expressed by the electron absorption
potential $V_{i}=Im\Sigma $. Such $E({\bf k})$ is smoothed in $k_{\bot }$
(the dispersions in ${\bf k}_{\parallel }$ remain nevertheless unsmoothed by
virtue of the surface-parallel invariance of the VLEED process) compared to
the quasiparticle $E({\bf k})$ {\em in the bulk}.\cite{CFS,Exi} The CPs
appear as the points of extremal (inverse) curvature of the the $k_{\bot }$
dispersions, and are intrinsically shifted from the corresponding CPs in the
bulk $E({\bf k})$, recovered in the $V_{i}=0$ limit, within a few tenths of
eV. To improve the accuracy of $\Delta \Sigma $ evaluation, based on
comparison with the DFT implying $V_{i}=0$, we corrected our VLEED data from
the $V_{i}$-induced shifts using a model calculation.

The model calculation employed the empirical pseudopotential method
including $V_{i}$. $E({\bf k})$ and $\phi _{{\bf k}}$ with complex $k_{\bot
} $ values were obtained by solving the Schr\"{o}dinger-like equation 
\[
\left( -\frac{\hslash ^{2}}{2m}\nabla ^{2}+V_{ps}-iV_{i}\right) \phi _{{\bf k%
}}=E\phi _{{\bf k}} 
\]%
where $V_{ps}$ is the pseudopotential. The VLEED spectra were calculated
within the matching approach (see details in [\ref{Exi}]). In principle,
matching calculations directly on top of the DFT calculations would be more
straightforward, but this would require more complicated techniques such as
the ${\bf k\cdot p}$ expansion.\cite{NbSe2} In addition to the standard
matching formalism, we included the surface barrier as an additional layer
with a skewed-cosine like potential and performed matching on the
vacuum-barrier and barrier-crystal planes. Compared to the step-like
barrier, this improved the VLEED spectral structure energies by $\sim 0.3\
eV $. The shifts between the $V_{i}=0$ CPs and $dT/dE$ extrema, found in the
model calculations, were then used to correct the experimental points back
to the $V_{i}=0$ limit to reflect the bulk $E({\bf k})$. This procedure also
compensated the slight shifts of the VLEED spectral structures caused by the
surface barrier and overlap of spectral structures.

The obtained unoccupied quasiparticle $E({\bf k})$ is shown in Fig.1
compared to the DFT and $GW$ calculations (see below). The experimental
points reveal only the bands with significant coupling to vacuum, and others
are not seen. We omitted the points where strong multiple-band hybridization
effects made it difficult to identify the CPs on the $\Gamma X$ line
reliably. The lower $X_{1}$-point measured by VLEED on the Cu(100) surface%
\cite{StrocovCu100} is also added.

Here we will also scrutinize our previous results on the valence band $E(%
{\bf k})$ from [\ref{CFS}]. They are perfectly consistent with the whole
body of previous experimental data compiled in [\ref{Courths84}], but have a
superior accuracy by virtue of employing a combined VLEED-PE method in which
VLEED\ was used to achieve full control of the 3-dimensional ${\bf k}$. The
experimental $E({\bf k})$ is shown in Fig.2 also compared with the
calculations. The original data from [\ref{CFS}] was processed to include
only the peaks relevant to the bulk states and compensate the lifetime
induced intrinsic shifts of the PE peaks notable in the bottom of the $sp$%
-band ($\sim 0.1eV$). In principle, the accuracy of these data can be
further improved by correcting for possible intrinsic shifts due to the
surface and matrix element effects, but this would require demanding
calculations within the one-step PE theory.\cite{Feibelman74}

\section{DFT and quasiparticle calculations}

Our DFT calculations used the Generalized Gradient Approximation (GGA)\cite%
{Perdew96} for $V_{XC}$. For Cu the results were however almost
indistinguishable from the LDA. A self-consistent full-potential linearized
augmented plane wave (FLAPW) method implemented in the {\it WIEN97} code\cite%
{Wien97} was used, with the basis extended by local orbitals to reduce the
linearization errors in the extended energy region of the unoccupied states.
Spin-orbit coupling was also included. It should be noted that nowadays all
state-of-art calculations on Cu agree within $\sim 100$ meV (at least in the
valence band) which implies that the true DFT $\varepsilon ({\bf k})$ is
achieved on this energy scale.The $\Delta \Sigma $ corrections to the DFT
eigenvalues, given by the expectation values $Re\int_{\Omega }\phi _{{\bf k}%
}^{\ast }\left( {\bf \Sigma }-V_{XC}\right) \phi _{{\bf k}}d{\bf r}^{\prime
} $ with the integration over the unit cell $\Omega $, were calculated in
the framework of the $GW$ approximation.\cite{Hedin69,Hedin99,Aryasetiawan98}
In this approximation the self-energy is given by \cite{Hedin69} 
\[
\Sigma ({\bf r},{\bf r}^{\prime };\omega )=\frac{i}{2\pi }\int d\omega
^{\prime }e^{i\eta \omega ^{\prime }}G({\bf r},{\bf r}^{\prime };\omega
+\omega ^{\prime })W({\bf r},{\bf r}^{\prime };\omega ^{\prime }) 
\]%
In practical calculations, the Green function $G$ is approximated by a
non-interacting one: 
\[
G({\bf r},{\bf r}^{\prime };\omega )=\sum_{{\bf k}}\frac{\psi _{{\bf k}}(%
{\bf r})\psi _{{\bf k}}^{\ast }({\bf r}^{\prime })}{\omega -\varepsilon _{%
{\bf k}}} 
\]%
In our case, $\{\psi _{{\bf k}},\varepsilon _{{\bf k}}\}$ are the LDA-DFT
wavefunctions and eigenvalues. The screened interaction $W$ is given by 
\[
W({\bf r},{\bf r}^{\prime };\omega )=\int d{\bf r}^{\prime \prime }v({\bf r}-%
{\bf r}^{\prime \prime })\epsilon ^{-1}({\bf r}^{\prime \prime },{\bf r}%
^{\prime };\omega ) 
\]%
where the dielectric matrix $\epsilon $ is calculated within the
random-phase approximation without employing the plasmon-pole approximation.
The present $GW$ calculations employed one-electron calculations using the
LDA exchange-correlation and the Linearized Muffin-Tin Orbitals - Atomic
Sphere Approximation (LMTO-ASA) method. In contrast to the pseudopotential
framework, this method allows to treat exchange-correlation with the core
states on the same footing as the valence states. Details of the computation
may be found in [\ref{Aryasetiawan98}]. The calculated $\Delta \Sigma $
values were added to the FLAPW eigenvalues whose computational accuracy at
higher energies is better. The $GW$ calculations on Cu are technically
rather demanding due to the presence of the $d$-states.

The results of the DFT\ and $GW$ calculations are also shown in Figs.1 and 2.

\section{Self-energy corrections}

The unoccupied quasiparticle $E({\bf k})$ in Fig.1 deviate from the DFT
results on the whole by $\sim 1\ eV$. However, the upper $\Delta _{1}$-band
displays an anomalously large shift by as much as $\sim 2.8\ eV$. Near $%
\Gamma $ the bands are shifted also very unevenly. These anomalies, missed
in the previous studies,\cite{Speier85} have been resolved here for the
first time by virtue of the single-state and ${\bf k}$-resolving nature of
VLEED. Reported previously,\cite{CFS,Marini02,Nilsson83} a peculiar
renormalization relative to the DFT$\;$is also observed in the valence band
in Fig.2, where the shifts of the $sp$-band (up to $+0.4\ eV$ in its bottom)
and the $d$-bands (about $-0.5\ eV$) even differ in sign. In the bottom of
the $d$-bands the deviations become smaller, which is due to hybridization
of the $d$- and $sp$-bands.\cite{Starnberg88} Through the unoccupied and
occupied states, the experimental shifts from the DFT are thus intriguingly
band- and ${\bf k}$-dependent.

The deviations can have two sources. Firstly, the static $V_{XC}$ is not
known exactly, and any DFT calculation has to resort to a reasonable
approximation such as the GGA. However, the small difference between LDA and
GGA results and the excellent description of ground-state properties suggest
that for Cu this problem is insignificant. The second, therefore
predominating, source is the excited-state self-energy effects due to the
difference of the dynamic exchange-correlation ${\bf \Sigma }$ from the
static one $V_{XC}$. Our comparison of the experimental quasiparticle and
DFT energies thus directly yields $\Delta \Sigma $. The $\Delta \Sigma $
values in the $X$-point, corresponding in our experiment\cite{CFS} to ${\bf K%
}_{\parallel }=0$, are given in Table 1 compared to the $GW$ calculations.

Inclusion of the dynamic exchange-correlation within the $GW$ approximation
are seen to vastly improve description of the experimental excitation
energies. In the unoccupied states the agreement is almost ideal, in
particular for the anomalous shift in upper $\Delta _{1}$-band, with the
only clear exception in the $\Gamma _{2^{\prime }}$ point. In the valence
band the agreement is also excellent for the $sp$-band. In the $d$-bands the
calculation correctly reproduces the sign of the $\Delta \Sigma $ shifts,
although underestimates their magnitude. Interestingly, previous
first-principles many-body calculations on Cu using formalisms other than $%
GW $ (see, e.g., [\ref{Zein95}]) failed to reproduce the $\Delta \Sigma $
behavior even qualitatively.

The observed self-energy anomalies in the unoccupied bands should be taken
into account in LEED surface crystallography (see, e.g., [\ref{VanHove79}]),
which so far relied on a monotonous energy dependence of $\Delta \Sigma $.
In our case, for example, the observed shifts of the VLEED spectral
structures would be interpreted as due to a surface relaxation, but they are
only due to the band- and ${\bf k}$-dependent self-energy behavior. The
observed excellent relevance of the $GW$ approximation for the unoccupied
states suggests to use it in LEED crystallography.

\section{Mechanisms of the self-energy behavior}

Although the observed self-energy behavior is well reproduced by the $GW$
calculations, their physical mechanisms remain obscured by the heavy
computational machinery. We will now endeavour to identify the possible
mechanisms, at least on a qualitative level.

\subsection{Local-density picture}

The observed anomalies in $\Delta \Sigma $ can be traced back to different 
{\em spatial localization} (SL) of the one-electron Bloch waves $\phi _{{\bf %
k}}$, i.e. the distribution of their weight within the unit cell. Such a
mechanism was originally suggested in [\ref{Nilsson83}]. It assumes the
local-density concept, i.e. neglects the non-locality of ${\bf \Sigma }$.

We start with a discussion of the $\Delta \Sigma $ behavior in the
homogeneous electron gas. The dynamic exchange-correlation $\Sigma $ as
function of momentum $k$ and electron density $n$ was calculated for this
model system by Hedin and Lundquist.\cite{Hedin71} From their data we
derived $\Delta \Sigma $ as the difference of $\Sigma $ to its value at the
Fermi wavevector $k_{F}$, which, by the DFT analogue to the Koopmans theorem,%
\cite{Koopmans} is essentially the static $V_{XC}$. The resulting
dependencies of $\Delta \Sigma $ on $E/E_{F}$ are shown in Fig.3 (left). The
crucial parameter to determine their behavior is the electron density.

In real crystals the electron density is inhomogeneous. The electron gas
plasmon dip in $\Delta \Sigma $ is damped (dashed line) by averaging over
the varying local density $n({\bf r})$ and by interband transitions. Most
importantly, as shown in Fig.3 (right), the effective electron density --
and consequently $\Delta \Sigma $ -- experienced by a one-electron
wavefunction depends now on its spatial localization: if $\phi _{{\bf k}}$
has large weight in the core region where $n({\bf r})$ is high, it will
experience a large-density energy dependence of $\Delta \Sigma $,
characterized by a strong $\Delta \Sigma $ repulsion from $E_{F}$. If $\phi
_{{\bf k}}$ expands well into the interstitial region with low $n({\bf r})$,
it will experience a small-density dependence with its smaller $\Delta
\Sigma $\ repulsion from $E_{F}$. Such an SL\ mechanism goes beyond the
usual wavefunction localization effect (see, e.g., [\ref{Jackson88}]) as not
only the strength of the localization (measured by the bandwidth) matters
but also the region of its localization.

In the valence band of Cu the SL mechanism has certainly some relevance (see
also [\ref{Nilsson83}]): The $d$-bands localized in the high-density core
region experience large effective density and, reflecting the $\Delta \Sigma 
$ behavior in Fig.3, shift lower in energy, whereas the $sp$-band with its
charge spread out into the low-density interstitial region experiences small
effective density and shifts in the opposite direction. A qualitative
support of this picture was obtained by calculating the effective densities $%
\langle \phi _{{\bf k}}|n({\bf r})|\phi _{{\bf k}}\rangle $ (see [\ref%
{GW-SRL}]), which well correlate with the experimental $\Delta \Sigma $
values both in the $d$- and $sp$-bands. Exclusion of the core states from
the total $n({\bf r})$ did not affect the correlation. However, the SL
mechanism alone can not provide any quantitative description of the
self-energy behavior mainly due to neglect of the non-locality of the ${\bf %
\Sigma }$ operator. Indeed, the corresponding calculation\cite{Nilsson83}
returned an unrealistic plasmon dip of $\Delta \Sigma $ in the unoccupied
bands and, although properly reproducing the sign of the $\Delta \Sigma $
shifts in the valence band, severely underestimated their magnitude.

In the unoccupied states, the SL\ mechanism could not be reconciled with the
experimental self-energy anomaly in the upper $\Delta _{1}$-band even
qualitatively unless we included into $n({\bf r})$ all core states down to
the 1$s$ state having a binding energy of $\sim 9$ keV. Any significant
effect of such a deep level on the dynamic exchange-correlation within the
local density picture seems unrealistic. This hints on an extremely
non-local mechanism of interaction with the core states, as we discuss below.

\subsection{$GW$ picture}

We will now analyze a connection between $\phi _{{\bf k}}$ and the
self-energy effects within the $GW$ framework, taking into account the
non-locality of the ${\bf \Sigma }$ operator acting as ${\bf \Sigma }\phi _{%
{\bf k}}=\int_{\Omega }{\bf \Sigma }\left( {\bf r},{\bf r}^{\prime }\right)
\phi _{{\bf k}}\left( {\bf r}^{\prime }\right) d{\bf r}^{\prime }$ . First,
we introduce a local quantity $\Delta \Sigma \left( {\bf r}\right) $
comprising local contributions to $\Delta \Sigma $, obtained by integrating
out the non-locality of ${\bf \Sigma }$ as 
\[
\Delta \Sigma \left( {\bf r}\right) =\phi _{{\bf k}}^{\ast }\int_{\Omega
}\left\{ {\bf \Sigma }\left( {\bf r},{\bf r}^{\prime }\right) -V_{XC}\left( 
{\bf r}\right) \delta \left( {\bf r}-{\bf r}^{\prime }\right) \right\} \phi
_{{\bf k}}\left( {\bf r}^{\prime }\right) d{\bf r}^{\prime } 
\]%
Integration of $\Delta \Sigma \left( {\bf r}\right) $ over the unit cell
with the weight $\propto r^{2}$ gives the total self-energy correction 
\[
\Delta \Sigma =4\pi \int_{\Omega }r^{2}\Delta \Sigma \left( {\bf r}\right) d%
{\bf r} 
\]

The calculated $\left| \phi _{{\bf k}}\right| ^{2}$, $\Delta \Sigma \left( 
{\bf r}\right) $ and $r^{2}$-weighted $\Delta \Sigma \left( {\bf r}\right) $
for the valence and unoccupied states in the $X$-point, obtained within the $%
GW$ framework, are shown in Fig.4. We note, firstly, that the total $\Delta
\Sigma $ is always formed by a balance between positive local $\Delta \Sigma
\left( {\bf r}\right) $ values in the core region and negative values in the
interstitial region (which is seen better in the $r^{2}\ast \Delta \Sigma
\left( {\bf r}\right) $ curves). Such a behavior of $\Delta \Sigma \left( 
{\bf r}\right) $ is in fact general. Indeed, if ${\bf \Sigma }$ is assumed
to be diagonal in the LDA Bloch basis as ${\bf \Sigma }\left( {\bf r},{\bf r}%
^{\prime }\right) =\sum_{{\bf k}}\phi _{{\bf k}}({\bf r})\Sigma _{{\bf k}%
}\phi _{{\bf k}}^{\ast }({\bf r}^{\prime })$, from the above equation we
obtain 
\[
\Delta \Sigma \left( {\bf r}\right) \sim \left( \Sigma _{{\bf k}%
}-V_{XC}\left( {\bf r}\right) \right) \left| \phi _{{\bf k}}\right| ^{2} 
\]%
In the core region (small ${\bf r}$) the negative $V_{XC}$ blows up, forcing 
$\Delta \Sigma \left( {\bf r}\right) $ to become large and positive in this
region. $\Sigma _{{\bf k}}$, determined by the particular $\phi _{{\bf k}}$,
acts as a constant negative offset. In the interstitial region (large ${\bf r%
}$) $V_{XC}$ decreases and $\Sigma _{{\bf k}}$ forces $\Delta \Sigma \left( 
{\bf r}\right) $ to become negative. In our case the off-diagonal elements
of ${\bf \Sigma }$ are small, and such a pattern holds well for all states
despite strong variations in the character of $\phi _{{\bf k}}$.

At first glance, the SL\ mechanism seems to revive in this picture, with an
amendment that the figure of merit is in fact the weight of $\phi _{{\bf k}}$
in the core region rather than that in the high-density region (these two
can somewhat differ). Indeed, if the weight of $\phi _{{\bf k}}$ shifts into
the core region, the positive contribution would increase and the
quasiparticle level would shift higher in energy. However, this is much
hampered by the $\Sigma _{{\bf k}}$ offset, which is also sensitive to the
character of $\phi _{{\bf k}}$ in a highly involved manner.

In view of the limitations of the SL\ mechanism, another mechanism of the
observed self-energy behavior can be suggested. Fig.4 shows that the valence
and two unoccupied $X_{1}$-states are different from others in that their $%
\phi _{{\bf k}}$ has a large $4s${\it \ }component, blowing up at the
nucleus (${\bf r}=0$). These -- and only these -- states experience an
anomalous {\em positive} $\Delta \Sigma $ shift on top of the general trend
that the $\Delta \Sigma $ repulsion from $E_{F}$ graually increases upon
going away from $E_{F}$.

To extend this observation to other points in the BZ, we scrutinized the
FLAPW calculations to determine the $4s$ projections (partial $4s$ charges)
of $\phi _{{\bf k}}$ inside the atomic spheres, which are given by
coefficients in the expansion $\phi _{{\bf k}}=\sum_{lm}A_{lm}u_{lm}\left( 
{\bf r}\right) Y_{lm}\left( \theta ,\phi \right) $. The calculated $4s$
projections are shown in Fig.5. Their comparison with the experimental
results in Figs.1 and 2 demonstrates that the anomalous positive $\Delta
\Sigma $ correspond everywhere to the $\Delta _{1}$-states having large $4s$
character. Moreover, in Fig.6 we show the experimental $\Delta \Sigma $
values as a function of energy and the $4s$ projections for the whole $%
\Gamma X$ line. Again, the anomalous positive $\Delta \Sigma $ on top of a
regular energy dependence correlate everywhere with the large $4s$ character
of $\phi _{{\bf k}}$, with the only clear exception in the $\Gamma
_{25^{\prime }}$-point.

The observed connection can not be explained within the above local picture,
because the $r^{2}$-weighting cancels the contribution of $\Delta \Sigma
\left( {\bf r}\right) $ from the nucleus region to the total $\Delta \Sigma $%
, see Fig.4. Moreover, other states such as $X_{5^{\prime }}$ and $%
X_{4^{\prime }}$, which have even more total weight in the high-density
region than the states with the $4s$ character, can nevertheless not
experience anomalous positive $\Delta \Sigma $. The observed connection can
only manifest a non-local exchange-correlation effect, acting through the $%
\Sigma _{{\bf k}}$ offset in the $\Delta \Sigma \left( {\bf r}\right) $
dependence.

In the following we propose a qualitative explanation for the above
anomalous behaviour. Let us focus ourselves on the states $X_{5^{\prime
}},X_{3}$ and $X_{1}$ which have almost the same correlation energy $\Sigma
_{C}$, as can be seen in Table 2. It implies that differences in the
self-energy shifts can only be due the differences in the exchange energy $%
\Sigma _{X}$ and $V_{XC}$. The trend in 
\[
\Sigma _{X}=-\sum_{{\bf q}}\int \phi _{{\bf k}}^{\ast }\left( {\bf r}\right)
\phi _{{\bf q}}^{\ast }\left( {\bf r}\right) v({\bf r}-{\bf r}^{\prime
})\phi _{{\bf q}}\left( {\bf r}^{\prime }\right) \phi _{{\bf k}}\left( {\bf r%
}^{\prime }\right) d{\bf r}d{\bf r}^{\prime } 
\]%
where ${\bf q}$ spans the occupied states, can be understood by analyzing
the character of the states. Indeed, the $X_{5^{\prime }}$-state, which has
a large $4p$ character, has the smallest $\Sigma _{X}$ among the three
states. This is because the occupied states have very little $4p$ character,
which makes the above exchange integral small. The $X_{3}$-state, on the
other hand, has a significant $3d$ character, which gives a large exchange
with the occupied $3d$ bands resulting in the largest $\Sigma _{X}$. The $%
X_{1}$-state, the anomalous one, has a large $4s$ character but the occupied 
$4s$ valence states are plane-wave like so that $\Sigma _{X}$ is somewhere
in between those of the $X_{5^{\prime }}$-state and $X_{3}$-state. The trend
in $V_{XC}$ seems clear from the charge distribution of the states, with the 
$X_{3}$-state being the most localized inside the muffin-tin sphere. The
physical picture arising from this qualitative analysis is that the $X_{1}$%
-state has an anomalously large shift because it has a large $4s$ character,
which makes its exchange relatively small, and at the same time its charge
distribution has a large weight inside the muffin-tin sphere where the
exchange-correlation potential is deep, which makes $V_{XC}$ large. These
two effects result in a large self-energy shift. We note that the amount of $%
4s$ character alone is probably insufficient to make a quantitative
connection to the self-energy correction. For example, the lower unoccupied $%
X_{1}$-state is less anomalous although it has a large $4s$ character too.
This can be understood from the fact that it has a significant $3d$
character (Fig.5), which increases the amount of exchange and reduces the
self-energy shift. Such an effect of the $3d$ character is consistent with
the negative $\Delta \Sigma $ shifts of the valence $3d$ bands. Moreover,
our qualitative picture neglects the exchange-correlation with the core
levels, while the recent $GW$ calculations\cite{Marini02} suggest that the
contribution of the $3s$ and $3p$ levels is significant.

The exchange contribution from the valence states, connected with the
wavefunction character, explains therefore the self-energy anomalies on top
of the SL mechanism. In the unoccupied states of Cu the exchange
contribution becomes critical.

\section{Is the $GW$ approximation accurate for the localized orbitals?}

The recent $GW$ calculation by Marini {\it et al},\cite{Marini02} performed
within the pseudopotential framework, have demonstrated a good agreement
with the experimental valence band of Cu both on the delocalized $sp$-band
and the localized $d$-bands. They conluded that the $GW$ approximation,
originally proposed to describe long-range charge oscillations, well extends
to the localized orbitals and short-range correlations found in Cu.

Our $GW$ results are somewhat different however. They also demonstrate
almost ideal agreement with the experiment on the delocalized bands such as
the valence $sp$-band, but on the localized $d$-bands they tend to
underestimate the negative $\Delta \Sigma $ shifts (see also Table 1).
Interestingly, both calculations yield a wrong sign of $\Delta \Sigma $ in
the bottom of the $d$-bands where they hybridize with the $sp$-band.
Compared to the pseudopotential framework, our LMTO-ASA calculations have
two fundamental advantages: 1) use of the true one-electron wavefunctions
rather than pseudowavefunctions; 2) explicit inclusion of all core states,
not only the $3s$ and $3p$ states as in [\ref{Marini02}]. Another recent $GW$
calculation based on the full-potential LMTO \cite{Kotani} also yielded
small $\Delta \Sigma $ shifts in the $d$-bands. The deviations from the
experiment, remaining in the $d$-bands, would therefore indicate certain
shortcomings of the $GW$ approximation in the description of localized
orbitals and short-range correlations. However, any unambiguous conclusions
on this point require further analysis, because the observed mismatch
between different $GW$ calculations is comparable with their numerical
accuracy.

\section{Other materials}

Experimental ${\bf k}$-resolved data on the self-energy effects in the
unoccupied bands of other materials is still scarce. In graphite\cite%
{Graphi3D} we have found only a small and regular $\Delta \Sigma $ shift to
higher energies (0.1 to 0.5 eV relative to the GGA-DFT) which however
notably increased above a distinct absorption threshold at 35 eV. A similar
situation has been found in NbSe$_{2}$.\cite{NbSe2} Such an increase is not
surprising, because the energy dependence of $Re{\bf \Sigma }$ is linked to
that of $Im{\bf \Sigma }$ via the Kramers-Kronig relations. Interestingly,
it was not reproduced by $GW$ calculations. In the valence band of graphite%
\cite{GraphiPE} we have identified the SL\ mechanism to cause different $%
\Delta \Sigma $ shifts of the $\sigma $- and $\pi $-states, having different
overlap with the in-plane oriented valence electron density. In [\ref%
{Shirley98},\ref{Heske99}] it was speculated that widening of the valence
band in graphite and other non-metals could result from electron
localization in high-density regions, in agreement with the SL\ mechanism.

The unoccupied bands of Ni\cite{Ni-VLEED} feature $\Delta \Sigma $ anomalies
resembling those of Cu. It seems that they are typical for noble metals. The
valence band of Ni (see, e.g., [\ref{Ni-PE}] and references therein) is
characterized by a narrowing of the $d$-bands ($\Delta \Sigma $ shifts
towards $E_{F}$) which is opposite to Cu. This self-energy effect has a
different origin - strong on-site correlations in the partially filled $d$%
-shell. The narrowing of the $sp$-band is however similar to Cu.
Interestingly, in [\ref{Starnberg88}] the ${\bf k}$-dependence of $\Delta
\Sigma $ in the whole valence band was decomposed into constant
contributions from the $sp$- and $d$-bands whose weight was determined only
by the band hybridization. Narrowing of delocalized bands has also been
observed for simple metals like Na.\cite{Shung87}

\section{Conclusion}

Excited-state self-energy effects in the unoccupied and valence band
electronic structure of Cu were investigated. The unoccupied quasiparticle
bands in the energy region up to 40 eV above $E_{F}$ are for the first time
mapped under full control of the 3-dimensional ${\bf k}$ using VLEED. They
demonstrate, similarly to the valence bands, a significant band- and ${\bf k}
$-dependence of the $\Delta \Sigma $ self-energy corrections to the DFT. The
observed $\Delta \Sigma $ behavior in the unoccupied and valence bands
correlates with the spatial localization of the one-electron wavefunctions
in the unit cell, and with their behavior in the core region expressed by
the $4s$ and $3d$ projections. These effects are described, correspondingly,
by the spatial localization mechanism based essentially on the local-density
picture, and by the non-local exchange interaction with the valence states.
In the unoccupied bands of Cu the latter is particularly important. Our $GW$%
{\it \ }quasiparticle calculations performed within the LMTO-ASA framework
yield almost ideal agreement with the experiment on the delocalized bands
such as the unoccupied and valence $sp$-bands, but on the localized $d$%
-bands they underestimate the $\Delta \Sigma $ shifts. This may indicate
certain limitations of the $GW$ approximation applied to the localized
orbitals and short-range correlations. The observed band- and ${\bf k}$%
-dependent self-energy effects in the unoccupied bands have an important
implication in LEED studies of the surface crystallography.

\section{Acknowledgements}

We thank R. Feder for valuable discussions. This work was supported by
Deutsche Forschungsgemeinschaft (grant Cl 124/5-1).

\begin{table}[tbp]
\caption{ The DFT eigenvalues in the $X$-point and the corresponding $\Delta
\Sigma $ corrections: experiment ($\Delta \Sigma _{EXP}$), our {\it GW}
calculations using LMTO-ASA ($\Delta \Sigma _{GW-LMTO}$), and {\it GW}
calculations using the pseudopotential scheme ($\Delta \Sigma _{GW-PP}$)
from [\ref{Marini02}]. Energies are in eV relative $E_{F}$. }
\label{Table1}\centering                                       
\begin{tabular}{cccccc}
& $E_{DFT-FLAPW}$ & $\Delta \Sigma _{EXP}$ & $\Delta \Sigma_{GW-LMTO}$ & $%
\Delta \Sigma _{GW-PP}$ &  \\ \hline
$X _{1}$ & 20.34 & 2.44 & 2.51 &  &  \\ 
$X _{3}$ & 17.63 & 0.45 & 0.43 &  &  \\ 
$X _{5^{\prime}}$ & 12.98/13.16 & 0.77/0.59 & 0.68 &  &  \\ 
$X _{1}$ & 7.13 & 0.54 & 0.81 &  &  \\ 
$X _{4^{\prime}}$ & 1.39 &  & 0.20 &  &  \\ 
$X _{5}$ & -1.55/-1.39 & -0.44/-0.60 & -0.26 & -0.64 &  \\ 
$X _{2}$ & -1.68 & -0.50 & -0.47 &  &  \\ 
$X _{3}$ & -4.51 & -0.34 & 0.05 & 0.10 &  \\ 
$X _{1}$ & -4.98 & -0.29 & 0.13 & 0.16 & 
\end{tabular}%
\end{table}

\begin{table}[tbp]
\caption{ The exchange and correlation energies for unoccupied bands in the $%
X$-point. All energies are in eV. }
\label{Table2}\centering                                     
\begin{tabular}{cccccc}
& $E_{DFT-FLAPW}$ & $\Sigma _{{\rm x}}$ & $\Sigma_{{\rm c}}$ & $V_{{\rm xc}}$
&  \\ \hline
$X _{1}$ & 20.34 & -10.07 & -5.89 & -19.83 &  \\ 
$X _{3}$ & 17.63 & -14.91 & -5.90 & -22.46 &  \\ 
$X _{5^{\prime}}$ & 12.98/13.16 & -6.64 & -5.68 & -14.19 &  \\ 
$X _{1}$ & 7.13 & -18.10 & -3.45 & -23.74 &  \\ 
$X _{4^{\prime}}$ & 1.39 & -10.48 & -2.18 & -13.92 & 
\end{tabular}%
\end{table}

\begin{figure}[tbp]
\caption{ VLEED quasiparticle unoccupied bands (dots) compared to the DFT $%
\protect\varepsilon ({\bf k})$ (solid lines). The deviations reveal the
band- and ${\bf k}$-dependent excited-state self-energy effects. The bands
with sufficient coupling to the vacuum are marked by gray shading. The
quasiparticle {\it GW } calculations are shown by open circles.}
\end{figure}

\begin{figure}[tbp]
\caption{ PE quasiparticle valence band compared to the DFT $\protect%
\varepsilon({\bf k}) $ and to the {\it GW } calculations similarly to Fig.1. 
}
\end{figure}

\begin{figure}[tbp]
\caption{ Mechanism of the spatial localization effect: ({\it left}): $%
\Delta \Sigma $ for the free electron gas. Its behavior depends on the
electron density. ({\it right}): In real crystals a wavefunction
concentrated in the core region experiences a high effective density,
whereas larger weight in the interstitial region (exaggerated, in reality
this case implies a more homogeneous distribution) leads to a lower density.
This results in different $\Delta \Sigma $ behavior.}
\end{figure}

\begin{figure}[tbp]
\caption{$\left| \protect\phi _{{\bf k}}\right| ^{2}$ and $GW$ calculated
local $\Delta \Sigma \left( {\bf r}\right) $ and $r^{2}$-weighted $\Delta
\Sigma \left( {\bf r}\right) $ for the unoccupied and valence states in the $%
X$-point. Note the change in the $r$ scale at 0.1 a.u. The curves are
normalized to set their variation to 1. $\Delta \Sigma \left( {\bf r}\right) 
$ is positive in the core region and negative in the interstitial region.
The $X_{1}$ states with the anomalous positive $\Delta \Sigma $ values have
the $4s$ character.}
\end{figure}

\begin{figure}[tbp]
\caption{Calculated $4s$ and $3d$ character of $\protect\phi _{{\bf k}}$,
expressed by the corresponding projections within the atomic spheres (in
electrons), and represented by the size of the circles.}
\end{figure}

\begin{figure}[tbp]
\caption{Experimental $\Delta \Sigma $ vs the DFT energy and $4s$ character
of $\protect\phi _{{\bf k}}$ from Fig.5, represented by the size of the
vertical bars. The dashed line is a guide for the eye showing a regular
energy dependence of $\Delta \Sigma $ for the states with negligible $4s$
character. The anomalous positive $\Delta \Sigma $ values correlate with the 
$4s$ character of the one-electron states.}
\end{figure}

\end{document}